# Direct wavefront sensing from extended objects: a general sensor

CHANGWEI LI,[1,2,3] YANTING LU,[1,2] AND SIJIONG ZHANG [1,2,3, *]

[1] *National Astronomical Observatories / Nanjing Institute of Astronomical Optics & Technology, Chinese Academy of Sciences, Nanjing 210042, China*
[2] *CAS Key Laboratory of Astronomical Optics & Technology, Nanjing Institute of Astronomical Optics & Technology, Nanjing 210042, China*
[3] *School of Astronomy and Space Science, University of Chinese Academy of Sciences, Beijing 100049, China*
*\* sjzhang@niaot.ac.cn*

**Abstract:** Wavefront sensing from an extended object is a challenging task since the phase to be sensed is disturbed by the phase generated from the structure of the extended object. To address this problem, a general wavefront sensor was proposed. The hardware of the sensor consists of a field lens, a collimating lens, a lenslet array, and a camera. The idea for its algorithm is to eliminate the phase caused by the extended object and reconstruct the point spread function through each lenslet. As a result, the scenario of wavefront sensing from an extended object has been converted to the conventional one from a point source. Numerical simulations and experiments both verify the feasibility and the accuracy of the proposed sensor.

## 1. Introduction

Wavefront sensing has played an important role in adaptive optics for the measurement of the distorted wavefront in many fields, such as astronomical imaging [1], retinal imaging [2], biological microscopy [3], and so on. Up to now, the widely used wavefront sensors, including the SHWFS [4], the curvature sensor [5], the shearing interferometer [6], and the pyramid wavefront sensor [7], can only sense the wavefront accurately from point sources. However, point sources are hardly to be found in the overwhelming majority of scenarios, and wavefront sensing has to rely on artificial ones. For example, in night astronomical imaging, natural bright stars can be served as guide stars for sensing the turbulence-induced wavefront distortion, but the sky coverage of the sufficiently bright stars is hopelessly low. On this account, astronomers are forced to utilize expensive artificial guide stars by shining a laser beam into the mesosphere [8]. In retinal imaging, a narrow laser beam is focused on the retina to generate a point source [9], which needs to be manipulated carefully to avoid permanent damage to the retina. In biological microscopy, the fluorescent beads are implanted into the samples to serve as the point sources [10]. In spite of aforementioned complexities in systems and manipulations, artificial point sources are still the best solution so far to these scenarios. But if the wavefront could be measured directly from the scene or object to be imaged, i.e., wavefront sensing from the extended objects, the heavy burdens of the generation of artificial point sources could be removed and the configuration of the adaptive optics systems could be consequently simplified. As a result, it is no doubt that wavefront sensing and further adaptive optics would have extensive applications.

The toughest issue in wavefront sensing from the extended object arises from the mixture of the phase caused by the propagation of light field from the extended object itself and the wavefront to be measured. For many years, researchers have been trying to crack this hard nut. However, to our knowledge, there is no any general solution for wavefront sensing from extended objects. Only some approaches have been developed for several application scenarios that require specific conditions. Name a few as follows. In solar adaptive optics [11], the SHWFS, equipped with the cross-correlation algorithm, has limited accuracy on the measurement of the atmospheric aberration due to the inappropriate usage of the same reference image for each subaperture. The partitioned aperture wavefront (PAW) [12] sensor, proposed for quantitative phase imaging, can only work on extended objects which are uniform and symmetrically distributed about the optical axis. Except for the above direct sensing methods, there are indirect wavefront measurement methods, such as phase retrieval [13] and phase diversity [14] approaches. These methods can, in principle, work for extended objects, but they have convergence and time-consuming problems in practical situations due to their dependency on iterative algorithms. It should be pointed out that this iterative scheme can also work in a hardware manner. For example, in a widefield microscopy application [15], wavefront sensing from extended objects is achieved by joint working of the PAW sensor, the science camera and the deformable mirror in an iteration manner. This alternative iteration scheme, de facto, belongs to the domain of indirect methods.

With the motivation of direct wavefront sensing from general extended objects in real time, a new wavefront sensor is proposed in this paper. The basic strategy of the proposed sensor is eliminating the influence of the extended object on the wavefront to be sensed. As a prerequisite to implement this strategy, a field lens is introduced into the sensor structure to ensure that subimages formed by the each lenslet are quasi-identical. With the quasi-identical subimages for each lenslet, the Fourier spectra of the subimages are quasi-identical, and then the elimination of the Fourier spectra of subimages of each lenslet can be performed on its pupil plane by division of the spectra of the subimages over the spectrum of the central subimage. After eliminating the Fourier spectra of subimages for each lenslet on its pupil plane, the corresponding point spread function can be extracted, and the scenario of wavefront sensing from extended objects has been converted to that from point sources. At last, the wavefront to be sensed can be reconstructed by the reconstruction method of the SHWFS. Since there is no limitation for the extended objects to be sensed from, we give this sensor a name as the general extended-object wavefront sensor (GEWFS) for the convenience of description thereafter.

This paper is organized as follows. The principles of the GEWFS wavefront sensor are presented in Section 2. The numerical simulation is given in Section 3. The experiments and results are shown in Section 4. Discussions are presented in Section 5, and conclusions are given in Section 6.

## 2. Principles

The basic idea of the GEWFS is converting the scenario of wavefront sensing from extended objects to that from point sources. Based on the theory of the isoplanatic image formation in Fourier optics, this goal is achieved by eliminating the Fourier spectrum of the subimage of the extended object for every lenslet on its pupil planes. The success of this elimination is guaranteed by the special structure of the GEWFS, which cannot be replaced by the classical SHWFS because of the different fields of view of its subapertures. The structure and the reconstruction method of the GEWFS will be described in details as following.

### 2.1 Structure of the GEWFS

The GEWFS, as shown in Fig. 1, is composed of a field lens, a collimating lens, a lenslet array, and a camera. Light from the extended object is converged at the center of the field lens, which is placed in the front focal plane of the collimating lens. After passing through the collimating lens, the collimated beam is incident on a lenslet array which is just behind the collimating lens. The camera, located at the rear focal plane of the lenslet array, records the subimage array formed by the lenslet array.

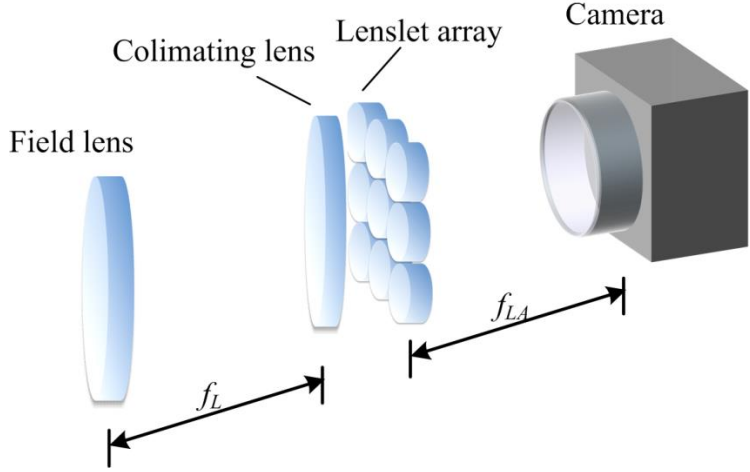

Fig. 1 Schematic of the GEWFS. $f_L$ and $f_{LA}$ are the focal lengths of the collimating lens and the lenslet array respectively.

The most important difference between the GEWFS and the SHWFS is that a field lens is located at the front focal plane of the collimating lens of the GEWFS. The usage of the field lens ensures that the subimages formed by every lenslet are quasi-identical. This point guarantees that the elimination of the Fourier spectrum of the subimages of the extended objects can be achieved using the method presented below.

## 2.2 Reconstruction method of the GEWFS

In the reconstruction method of the GEWFS, the most difficult point is how to eliminate the information of the extend object which will disturb the wavefront reconstruction. Here, the theoretical basis of our solution is the theory of image formation described by Fourier optics. In the following, firstly, the principles of the elimination of the information of the extended object and the reconstruction of the point spread functions of each lenslet are presented. Then, steps for wavefront reconstruction of the GEWFS are given.

To present how to eliminate the information of the extended object, the demonstration over one subaperture is taken as an example. According to the Fourier optics, for each lenslet, its subimage can be expressed as the convolution of the ideal image of the object and the point spread function of this lenslet

$$i_{m,n}(x,y) = o(x,y) \otimes p_{m,n}(x,y) \tag{1}$$

where $i_{m,n}(x,y)$ stands for the subimage, $o(x,y)$ the ideal image of the object, $p_{m,n}(x,y)$ the point spread function, and the subscripts $m$ and $n$ are the row and column indexes of the corresponding subaperture. Note that the point spread function, $p_{m,n}(x,y)$, which contains the local wavefront, can be used to reconstruct the wavefront. Applying the convolution theorem of Fourier transform to Eq. (1), the Fourier spectrum of the subimage can be written as

$$I_{m,n}(f_x,f_y) = O(f_x,f_y) T_{m,n}(f_x,f_y) \tag{2}$$

where $I_{m,n}(f_x,f_y)$ is Fourier transform of the subimage $i_{m,n}(x,y)$, $O(f_x,f_y)$ the Fourier spectrum of $o(x,y)$, $T_{m,n}(f_x,f_y)$ the optical transfer function of the corresponding lenslet. Actually, the amplitude of $T_{m,n}(f_x,f_y)$ is the autocorrelation of the pupil function of the corresponding lenslet, and the phase of $T_{m,n}(f_x,f_y)$ is the local wavefront fell on that lenslet.

Then, the key point of wavefront reconstruction is the elimination of $O(f_x, f_y)$ in Eq. (2). Due to the usage of the field lens, the subimages obtained through all subapertures are quasi-identical, and so are their Fourier spectra. So it is very convenient to eliminate the Fourier spectrum of the ideal image of the object by dividing the Fourier spectrum $I_{m,n}(f_x, f_y)$ with the Fourier spectrum of a reference subimage. In this paper, the subimage of the central subaperture, is selected as the reference one. If the Fourier spectrum of the reference subimage is denoted as $I_r(f_x, f_y)$, the optical transfer function, whose phase angle is relative to that of the reference one, of the lenslet numbered $(m,n)$ will be reconstructed by

$$T'_{m,n} = |T_{m,n}| \exp[i(\varphi_{m,n} - \varphi_r)] = (P \star P) \exp\left[i \cdot \arg\left(\frac{I_{m,n}}{I_r}\right)\right] \quad (3)$$

where $\arg(\cdot)$ stands for the extraction of the phase angle, $\star$ the operation of correlation, $P$ the pupil function of the lenslet, $\varphi_{m,n}$ the local wavefront fell on the subaperture numbered $(m,n)$, and $\varphi_r$ the local wavefront fell on the reference subaperture. For brevity, the independent variables in eq. (3) are omitted. In Eq. (3), since the operation of $\frac{I_{m,n}}{I_r}$ is sensitive to noises, in practice, a small quantity is added on $I_r$ to suppress the influence of the noise on extracting $T'_{m,n}$ which in fact is a way of Wiener filtering.

In Eq. (3), there are two assumptions. First, the pupil function of each lenslet can be regarded to be identical, which is basically satisfied due to the precision manufacturing of the lenslet array. Second, the local wavefront fell on each subaperture can be approximated as a plane. If this assumption dose not satisfied, the local wavefront fell on the reference subaperture will be mixed into the reconstructed optical transfer functions, and will lead to the inaccuracy of the reconstructed wavefront. Therefore, the ratio of the diameter of the input beam and that of the lenslet must be large enough to ensure that the local wavefront fell on each subaperture can be approximated as planes.

Note that, the sum of the local wavefront and the phase of $O(f_x, f_y)$ maybe exceed $2\pi$, and the phase will be wrapped. Even in this situation, the Fourier spectrum of the object can still be eliminated by the procedure indicated in Eq. (3). In Eq. (3), the extracted phase angle is $\mod\left[(\varphi_0 + \varphi_{m,n}) - (\varphi_0 + \varphi_r), 2\pi\right]$, where $\varphi_0$ is the phase angle of $O(f_x, f_y)$. Although the extracted phase is indeed not equal to the phase angle $(\varphi_0 + \varphi_{m,n}) - (\varphi_0 + \varphi_r)$, the only difference between them is an integral multiple of $2\pi$ at each location with a phase jump. So the two fields, $|T_{m,n}| \exp\{i \mod[(\varphi_0 + \varphi_{m,n}) - (\varphi_0 + \varphi_r), 2\pi]\}$ and $|T_{m,n}| \exp[i(\varphi_0 + \varphi_{m,n}) - (\varphi_0 + \varphi_r)]$, will be identical. This means that the information of the extended object can still be eliminated when the sum of the local wavefront and the phase of $O(f_x, f_y)$ exceeds $2\pi$.

Furthermore, by performing the inverse Fourier transform on the reconstructed optical transfer function, $T'_{m,n}$, the corresponding point spread function, $p'_{m,n}$, can be acquired. According to the shift theorem of Fourier transform, the phase angle of $p'_{m,n}$, i.e., the local wavefront, in frequency domain, corresponds to a spatial displacement of the reconstructed point spread function, $p'_{m,n}$, in spatial domain. In the same way, the reconstructed point spread functions over all lenslets can be obtained. By now, the problem of wavefront sensing from extended objects has been converted to that from point sources, and the input wavefront can be easily reconstructed by the reconstruction algorithm of the SHWFS [16].

According to the principles of the GEWFS presented above, its reconstruction method can be divided into five steps:

Step 1 Extract the subimage for each subaperture from the subimage array recorded by the camera of the GEWFS respectively;

Step 2 Perform the Fourier transform on each subimage to acquire the Fourier spectrum on the pupil plane of each lenslet;

Step 3 Eliminate the Fourier spectrum of the subimage on the pupil plane of each lenslet, and reconstruct its optical transfer function;

Step 4 Acquire the point spread function for each lenslet by performing inverse Fourier transform on the corresponding reconstructed optical transfer function;

Step 5 Extract the shift of centroid of the point spread function for each lenslet, and reconstruct the input wavefront by the algorithm of the SHWFS.

## 3. Numerical simulations

In this section, the performances of the reconstruction method of the GEWFS are validated using the simulated data. The pitch and the focal length of the lenslet array used in simulation are 150 µm and 3.72 mm respectively. The pixel number of the detector is 256×256 with a pixel size of 9.9×9.9 µm. Letter "F" is used as the extended object to exam the proposed reconstruction method because of its asymmetric structure. The wavefronts, a defocus (labeled as wavefront A hereafter) and a one generated by a linear combination of Zernike polynomials (labeled as wavefront B hereafter), are used as input distortions to generate the simulated intensity distributions, which are simulated by the propagation of angular spectrum. In last step of our reconstruction method, the wavefronts are reconstructed by Zernike polynomials from the 4rd to 36th terms.

### 3.1 Reconstruction of wavefront A

As shown in Fig. 2(a), wavefront A, with a root mean square (RMS) value of 0.22 µm and a peak to valley value (PV) of 0.75 µm, is used as the input wavefront, and the corresponding simulated intensity distribution is shown in Fig. 2(b).

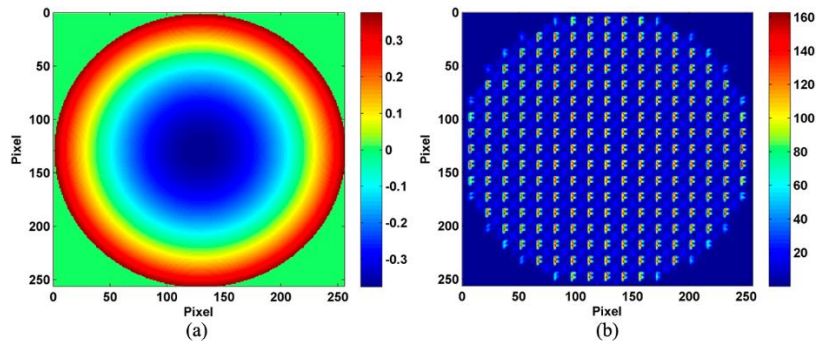

Fig. 2. Wavefront A (a), scales in µm, and the corresponding simulated intensity distribution (b).

Figure 3 shows the reconstructed wavefront and the reconstruction error of wavefront A. The RMS and the PV values of the residual wavefront error are 0.027 µm and 0.15 µm, respectively.

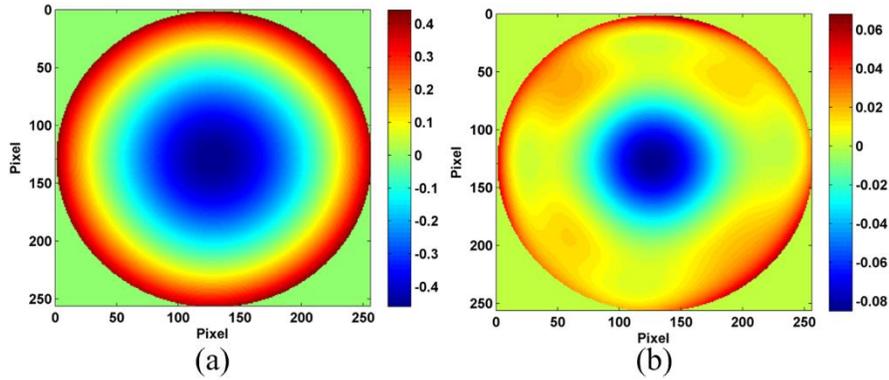

Fig. 3. Reconstructed wavefront (a) and the reconstruction error (b) of wavefront A, scales in µm.

## 3.2 Reconstruction of wavefront B

Wavefront B, a linear combination of Zernike polynomials from the 4rd to 15th terms with random coefficients, is shown in Fig. 4(a). The RMS and the PV values of the wavefront are 0.19 µm and 1.38 µm, respectively. The corresponding simulated intensity distribution is shown in Fig. 4(b).

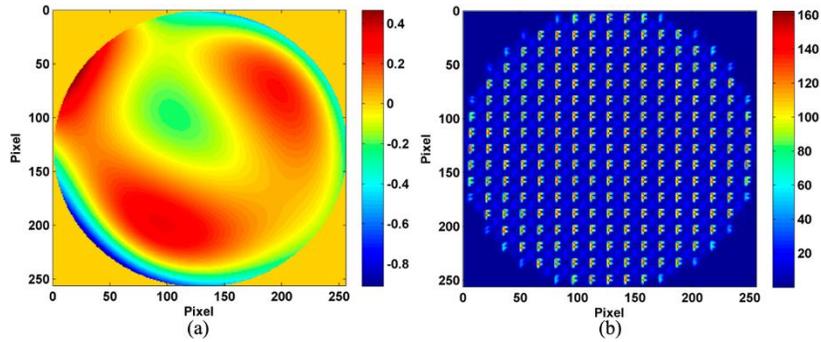

Fig. 4. Wavefront B (a), scales in µm, and the corresponding simulated intensity distribution (b).

Figure 5 shows the reconstructed wavefront and the reconstruction error of wavefront B. The RMS and the PV values of the residual wavefront error are 0.015 µm and 0.082 µm, respectively. The Zernike coefficients of the input (red) and reconstructed (blue) wavefronts are shown in Fig. 6.

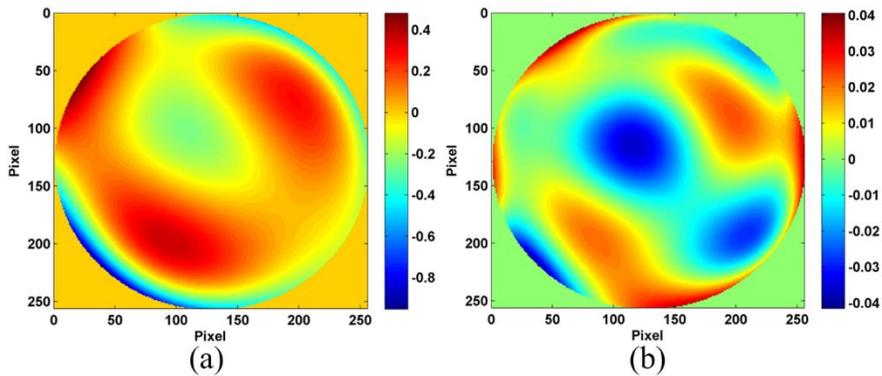

Fig. 5. Reconstructed wavefront (a) and the reconstruction error (b) of wavefront B, scales in µm.

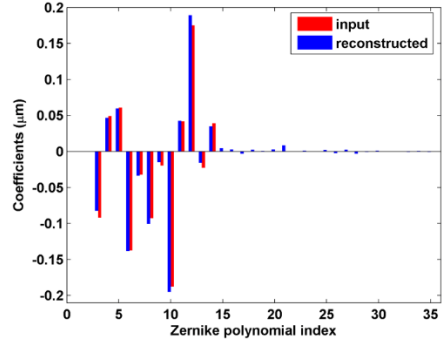

Fig. 6. Zernike coefficients of the input (red) and reconstructed (blue) wavefront B, scales in µm.

The results of the numerical simulation show that wavefront sensing from the extended object using the proposed reconstruction method is accurate and feasible. In the following section, experiments will be performed for further verification.

## 4. Experiments

### 4.1 Experimental setup

Figure 7 shows the optical setup for performing the experiments of wavefront sensing from extended objects. The light beam, from a broadband halogen fiber optic illuminator (Torlabs, OSL1), is homogenized by a piece of frosted glass. Illuminated by this homogenized light beam, a small portion of a USAF 1951 resolution test target (Edmund, 3" × 3" positive) is used as the extended object; besides, a point source, used for testing the accuracy of the GEWFS, is obtained by placing a pinhole (Edmund, 20 µm) between the frosted glass and the test target. The beam coming from an extended object (or a point source) is first collimated by a lens, L1. Then, the collimated beam is disturbed by a piece of spherical aberration compensation plate (Edmund, +1.00 λ aberration at the wavelength of 0.587 µm). The disturbed beam is converged by lens L2 and then incident on the GEWFS. The GEWFS, as shown in the red dashed box in Fig. 7, is consisted of a field lens, a collimating lens, a lenslet array (150 µm pitch, 3.72 mm focal length), and a CCD camera (640 × 480 pixels, 9.9 × 9.9 µm pixel size). All components of the GEWFS are off-the-shelf products, so this sensor is easy to build.

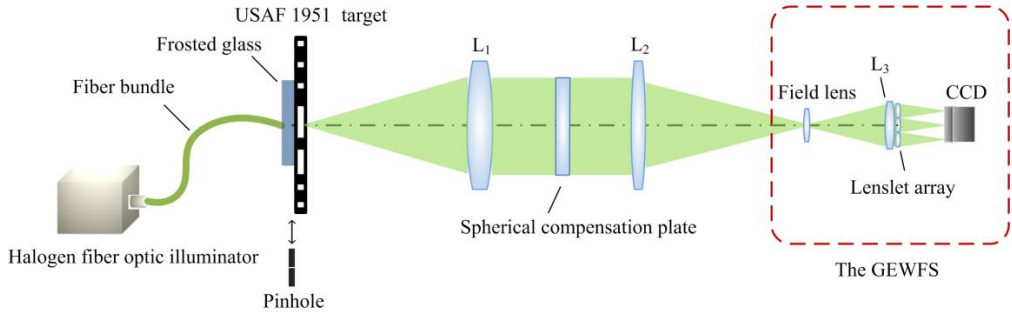

Fig. 7. Schematic of the experiment setup for wavefront sensing by the GEWFS (in the red dashed box).

### 4.2 Experimental results

In experiments, firstly the aberration, introduced by the spherical compensation plate, is measured by a SHWFS from a point source generated by inserting a pinhole between the frosted glass and the test target. Secondly, the same aberration is measured by the GEWFS from different extended objects. And finally, the aberrations sensed from different extended objects are carefully evaluated by that sensed from the point source.

Figure 8 shows the intensity distributions recorded by the sensor for a point source (P) and three different extended objects (E1, E2, and E3), respectively. The wavefronts, respectively, reconstructed from sources P, E1, E2, and E3, are shown in Fig. 9. Obviously, the reconstructed wavefronts from different extended objects appear the same as that from point sources, except a little bit differences in the central part of the wavefronts. Figure 10 shows the Zernike coefficients of the reconstructed wavefronts, respectively, from the point source and the three different extended objects. Since the wavefront measured is introduced by a spherical aberration compensation plate in the experimental setup, the components in the reconstructed wavefronts are the defocus and the spherical aberrations (corresponding to the fourth and eleventh terms of Zernike polynomials). The Zernike coefficients, reconstructed from sources E1, E2, and E3, agree well with those reconstructed from source P, and the biggest differences in Zernike coefficients, respectively, between E1, E2, E3 and P, are 0.023 μm, 0.024 μm, 0.014 μm.

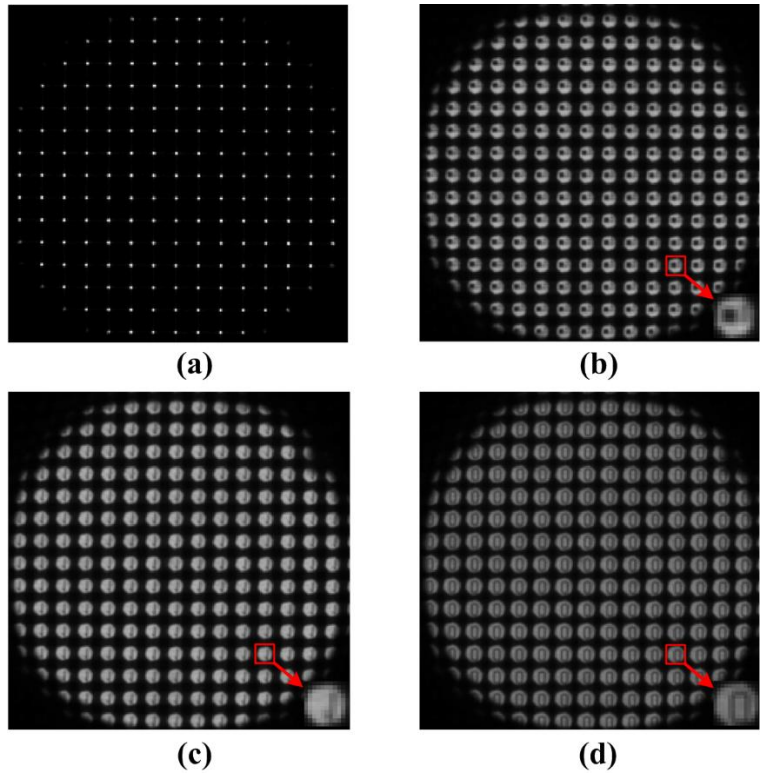

Fig. 8. Subimage arrays of sources P (a), E1 (b), E2 (c), and E3 (d)

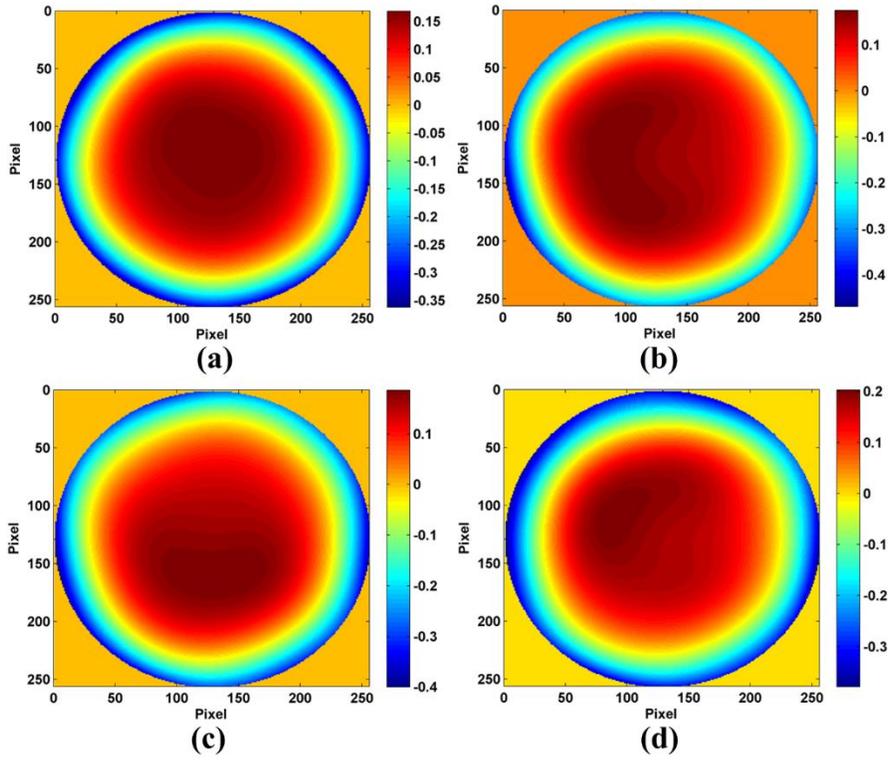

Fig. 9. Reconstructed wavefronts, respectively, from P, E1, E2, and E3, scales in µm

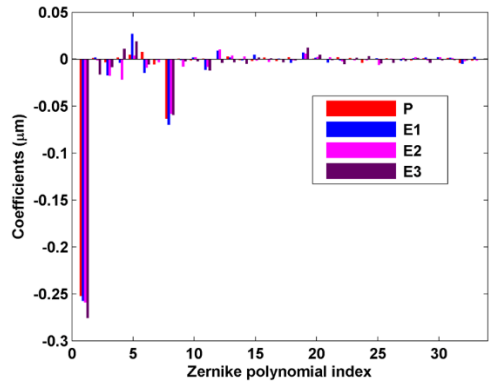

Fig. 10. Zernike coefficients of the reconstructed wavefronts, respectively, from P, E1, E2, and E3, scales in µm.

Table 1 lists the RMS and the PV errors of the differences between the reconstructed wavefronts from the three different extended objects and that from the point source, respectively. From the RMS and PV errors shown Table 1, it can be seen that the wavefronts reconstructed from the three different extended objects agree well with that reconstructed from the point source.

**Table 1. RMS and PV errors of the differences between the wavefronts reconstructed from E1, E2, E3, and P**

| Differences between wavefronts reconstructed from | RMS errors (in $\mu$m) | PV errors (in $\mu$m) |
|---|---|---|
| E1 and P | 0.024 | 0.20 |
| E2 and P | 0.023 | 0.17 |
| E3 and P | 0.026 | 0.17 |

Finally, The GEWFS is evaluated by the statistical hypothesis testing. The aberration introduced by one piece of spherical aberration compensation plate is measured by SHWFS from a point source and by our sensor from an extended object respectively, and the wavefront sensing experiments from point source/extended objects are repeated 30 times, respectively. Two two-sample *t*-tests [17] are performed, respectively, on the defocus and the spherical coefficients of results from the SHWFS and the GEWFS, since the defocus and the spherical aberration are the two only components of the aberration introduced by the spherical aberration compensation plate. The results of these two two-sample *t*-tests show that the results of the defocus and spherical coefficients from the GEWFS have no significant difference at 5% level compared with those from SHWFS, which means that the GEWFS is blind to the disturbance caused by extended objects. And this blindness is just the characteristic required for wavefront sensing from extended objects.

Taken together, these results all indicates that the proposed algorithm is practically feasible for wavefront sensing from extended objects. Beside, since there is no special requirement on the extended objects in our experiments, the GEWFS is a general sensor for wavefront sensing from extended objects.

## 5. Discussions

In this section, several characteristics of the GEWFS will be discussed.

**Accuracy**. Note that, in order to ensure the accuracy of the proposed algorithm, the ratio of the diameter of the input beam and that of the lenslet should be large enough. As mentioned in step 3 in Section 2, if this condition does not satisfied, the reconstructed wavefront may be inaccurate.

**Dynamic Range**. For each subaperture, the subimage of the extended object occupies a certain area on the corresponding region of the imaging plane, such that the dynamic range of sensing from extended objects will be smaller than that from a point source in the same condition.

**Real-time capability**. Since the Fourier transforms is performed on each subimage, the computational complexity of the proposed algorithm is $O(NM \log M)$, where $N$ is the number of subapertures used for wavefront reconstruction, and $M$ the number of pixels of the subimage. Since the Fourier transforms are performed on the subimages, the size of which is quite small relative to that of the whole image recorded by the CCD camera of the GEWFS, the computational burden of our reconstruction method is not heavy. Therefore, the GEWFS can work with extended objects in real time.

## 6. Conclusions

The GEWFS for direct wavefront sensing from extended objects has been presented. A field lens employed in the GEWFS to ensure that all subimages formed by each lenslet are quasi-identical. With this guarantee, the Fourier spectra of the extended objects are eliminated on the pupil plane of each lenslet by division of the spectra of the subimages over the spectrum of the central subimage. Then, the optical transfer function through each lenslet and corresponding point spread function can be reconstructed. In such a way, the problem of wavefront sensing from extended objects has been converted into that from point sources, which can be easily solved. Numerical simulations show the effectiveness of the GEWFS. The experiments for comparisons of the GEWFS and the SHWFS show that the wavefront reconstructed from extended objects are in good agreement with that reconstructed from the point source, indicating that the GEWFS is accurate and feasible for real-time wavefront sensing. In our opinion, the GEWFS overcomes the limitation of wavefront sensing by means of point sources, and will broaden the fields of application of adaptive optics.

**Funding**

National Natural Science Foundation of China (Grant Nos. 11873069 and11703060)

**Acknowledgements**

This work was supported by the grants of the National Natural Science Foundation of China (Grant Nos. 11873069 and11703060).